\documentclass
[10pt,superscriptaddress,secnumarabic,amssymb,amsmath,nobibnotes,aps,prd,showkeys,showpacs,nofootinbib,onecolumn,notitlepage]{revtex4-1}
\usepackage{setspace}
\usepackage{color}
\usepackage{amsmath}
\usepackage{amsfonts}
\usepackage{verbatim}
\usepackage{amssymb}
\usepackage{graphicx,bm}
\usepackage{graphicx}
\usepackage{graphicx}
\usepackage{epstopdf}
\usepackage{epsf}
\usepackage[caption=false]{subfig}
\definecolor{darkgreen}{rgb}{0,0.35,0}

\providecommand{\U}[1]{\protect\rule{.1in}{.1in}}
\begin{document}

\title{Integrability of the mixmaster model}
\author{N. Dimakis}
\email{nsdimakis@scu.edu.cn}
\email{nsdimakis@gmail.com}
\affiliation{Center for Theoretical Physics, College of Physical Science and Technology Sichuan University, Chengdu 6100064, China}
\author{Petros A. Terzis}
\email{pterzis@phys.uoa.gr}
\affiliation{Nuclear and Particle Physics Section, Physics Department, National and Kapodistrian University of Athens, GR 157–71 Athens, Greece}
\author{T. Christodoulakis}
\email{tchris@phys.uoa.gr}
\affiliation{Nuclear and Particle Physics Section, Physics Department, National and Kapodistrian University of Athens, GR 157–71 Athens, Greece}


\begin{abstract}
The mixmaster model has always been a field of controversy in the literature regarding its (non)integrability. In this work, we make use of a generalized definition of a class of nonlocal conserved charges in phase space to demonstrate that the anisotropic Bianchi type IX model in vacuum is -at least locally - Liouville integrable, thus supporting the findings of previous works pointing to this result. These additional integrals of motion that we use can be defined only due to the parametrization invariance of the system and can be seen to possess an explicit dependence on time. By promoting the time variable to a degree of freedom, we demonstrate the existence of two sets of four independent conserved charges that are in involution, thus leading to the characterization of the system as integrable in terms of the Liouville-Arnold theorem.
\end{abstract}

\maketitle

\section{Introduction}

The general Bianchi type IX cosmological model has always been a source of great debate in the literature, both in regards to its integrability \cite{Cushman,Intp1,Cotsakis,Intp2,Intp3,Intp4,Intp5} and its chaotic nature \cite{Barrow, chaos1,chaos3,chaos2,Matsas2,Ellis,Rugh, Hobill,Berger1,Matsas1,Pullin,chaos4.0,chaos4,chaos5,chaos6,chaos7,Motter}. The situation gets more complicated given the fact that different notions of integrability are used that are not necessarily equivalent (e.g., Liouville \cite{Arnold} or Painlev\'e \cite{Painleve} integrability). Even worse, in many cases gauge dependent methods are being applied to a parametrization invariant system, leading to conflicting results.

A general review of the basic features in spatially homogeneous cosmologies can be found in \cite{Ryan,Jan}. The study and importance of the Bianchi type IX geometry in gravitational physics began with the pioneering work of Misner and his mixmaster model \cite{Misner}. With the works of Belinskii, Khalatnikov and Lifshitz \cite{BKL1,BKL1b,BKL2} it has been shown that the system evolves through successive Kasner epochs and near the singularity adopts an oscillatory behaviour. They have also suggested that the anisotropic Bianchi type IX may be additionally used to describe more general gravitational solutions. Several numerical studies for the Bianchi IX have been performed in support of these results \cite{Ryan0,Berger2,Berger3}. A first important theorem on the asymptotic behaviour towards the singularity was derived in \cite{Ringstrom} and further discussed in \cite{Uggla}. This was made possible with the use of variables that were previously introduced in \cite{WHsu} for class A Bianchi cosmological systems. The latter have served for the asymptotic dynamical analysis in other Bianchi cosmologies as well \cite{LeBlanc1,Bergerf,LeBlanc2}.

To make a quick resume of the debate so far, we mention that in \cite{Intp1,Cotsakis} it was claimed that the Bianchi type IX dynamical system passes the Painlev\'e test. Later in \cite{Intp2} it was shown that for negative values of the energy the system does not pass the test, thus criticising the approach taken in \cite{Intp1,Cotsakis}, which made no such distinction. The authors of \cite{Intp1} revisited their study in \cite{Intp3}, and concluded that the matter of the Painlev\'e integrability of the model remains open. In \cite{Intp4}, it was claimed that the Bianchi type IX does not pass the Painlev\'e test, while on the other hand, the authors of \cite{Intp5} argue that the issues raised in \cite{Intp2,Intp3,Intp4} cannot exclude integrability. At the same time, a discussion was raised regarding the possible chaotic behaviour of the model in terms of the Lyapunov exponents and their non-invariance under the adoption of different time gauges. Initially, it was shown in \cite{Barrow} with the use of the Gauss map, that one is led to a positive Lyapunov exponent. However, later analysis \cite{Matsas2,Ellis} implied that the Lyapunov exponents tend to zero as the system asymptotically approaches the singularity. The discrepancy was traced to the adoption of different time gauge choices in the relevant studies \cite{Hobill,Berger1,Matsas1}. For a review on this matter see \cite{Burdrev} and references therein. In \cite{Hobill} there was also stated  one of the basic problems in the analysis about the emergence of chaos in the Bianchi type IX vacuum model. Namely, the satisfaction of the constraint equation whenever numerics are involved or the precise conditions under which the Kasner Bianchi type I solution is assumed at an approximate regime.

In regards to exact methods, works like \cite{non1,non2} have excluded the existence of first integrals which are analytic functions of the phase space variables for the general Bianchi type IX system that is, apart from the Hamiltonian itself. However, in the usual studies of integrability, one usually considers as possible integrals of motion, quantities that are functions just of the position and the momenta, i.e. $Q\equiv Q(q,p)$. In this work, we adopt a different approach: We demonstrate that there exist conserved charges that possess an explicit dependence in time $Q\equiv Q(t,q,p)$ in the form of a nonlocal term. That is a term which is given as an integral over time of phase space functions. These nonlocal conserved charges appear in parametrization invariant systems and are not a consequence of Noether's theorem \cite{tchris}, i.e. they do not result from variational symmetries of the action; but rather, they are a generalization of Kucha\v{r}'s conditional symmetries \cite{Kuchar}. Their existence is tied to the parametrization invariance of the system and thus, it is important to consider the latter in its most general form avoiding any gauge fixing prior to the derivation of the symmetries.

The structure of the paper is as follows: In section \ref{sec2}, we begin by writing the equivalent minisuperspace system that reproduces Einstein's equation in vacuum for the mixmaster model. Next, in section \ref{sec3} we introduce the Hamiltonian description of the model and obtain all those conserved charges that are at most linear in the momenta and possess an additional nonlocal part. We subsequently fix the gauge and treat the time variable as an additional degree of freedom. The latter is done in order to treat the explicit time dependence of these integrals in phase space. We prove that there exists the necessary number of independent commuting integrals of motion so that the system can be characterized as locally (Liouville) integrable. Something which is in conjunction with the result of \cite{Cushman} where the local integrability of the system was studied from a different perspective. Before we conclude with our final remarks, we remind in section \ref{sec4} a few facts from the theory of singular systems that we consider to play a significant role in our proof and its relation to existing results.

\section{The equivalent minisuperspace system} \label{sec2}

The mixmaster universe is described by the line element
\begin{equation}  \label{lineel}
ds^2 = - N(t)^2 dt^2 + \gamma_{\alpha\beta}(t) \sigma^\alpha_i (x)
\sigma^\beta_j(x) dx^i dx^j, \quad i,j,\alpha,\beta = 1,2,3 ,
\end{equation}
where $x^i =(x,y,z)$ are the spatial coordinates, $\gamma_{\alpha\beta}=\mathrm{diag}(a(t)^2,b(t)^2,c(t)^2)$ is the scale factor matrix and the $\sigma^\alpha_i$ are the 1-forms corresponding to the invariant basis associated to the three dimensional group of isometries acting simply transitively on the spatial surface $t=$constant:
\begin{equation}
  \begin{split}
    \sigma^1 &=   \sin x \cos z \mathrm{d}y - \sin z \mathrm{d}x \\
    \sigma^2 &=  \sin x \sin z \mathrm{d}y + \cos z \mathrm{d}x \\
     \sigma^3 & = \cos x \mathrm{d} y + \mathrm{d} z .
  \end{split}
\end{equation}

These 1-forms satisfy the well known Maurer-Cartan equations \cite{Ryan}
\begin{equation} \label{sigmarel}
d \sigma^\alpha = \frac{1}{2} C^\alpha_{\beta\gamma} \sigma^\beta \wedge
\sigma^\gamma,
\end{equation}
where in this case $C^\alpha_{\beta\gamma} = \epsilon_{\alpha\beta\gamma}$, with $\epsilon_{\alpha\beta\gamma}$ being the Levi-Civita symbol in three dimensions ($\epsilon_{123}=+1$), since the three dimensional group of isometries in this case is the rotation group $SO(3)$.

For the rest of our analysis - and in order to have a simplified minisuperspace metric - we assume the description of the system in the Misner variables \cite{Ryan}, where
\begin{equation}  \label{Misnvar}
a= e^{\beta_1 + \sqrt{3} \beta_2 - \Omega}, \quad b= e^{\beta_1 - \sqrt{3}
\beta_2 - \Omega}, \quad c = e^{-2 \beta_1 - \Omega} .
\end{equation}
If we substitute the line element \eqref{lineel} into the gravitational action $S=\int\sqrt{-g}R d^4x$, with $g=\mathrm{Det}g_{\mu\nu}$, and integrate out the nondynamical degrees of freedom, then we are left with the minisuperspace Lagrangian
\begin{equation}\label{Lag}
  L = \frac{1}{2N} G_{\alpha\beta}\dot{q}^\alpha \dot{q}^\beta -N V(q)
\end{equation}
where the dots denote differentiation with respect to the time variable $t$. The configuration space variables are $q=(\Omega,\beta_1,\beta_2)$ (as we said we work in the Misner variables \eqref{Misnvar}) and the potential $V(q)$ is
\begin{equation}
  \begin{split}
    V(\Omega,\beta_1,\beta_2)= &\frac{1}{2} e^{4 \beta_1-4 \sqrt{3} \beta_2-\Omega}-e^{-2 \beta_1-2 \sqrt{3} \beta_2-\Omega} +\frac{1}{2} e^{4 \beta_1+4 \sqrt{3} \beta_2-\Omega} +\frac{1}{2} e^{-8 \beta_1-\Omega}\\
    &-e^{4 \beta_1-\Omega } -e^{-2 \beta_1+2 \sqrt{3} \beta_2-\Omega} .
  \end{split}
\end{equation}
The minisuperspace metric $G_{\alpha\beta}$ is diagonal in these coordinates and reads
\begin{equation}\label{min}
  G_{\alpha\beta} = 12 e^{-3 \Omega} \begin{pmatrix}
                                       -1 & 0 & 0 \\
                                       0 & 1 & 0 \\
                                       0 & 0 & 1
                                     \end{pmatrix}.
\end{equation}
The minisuperspace Lagrangian \eqref{Lag} correctly reproduces the Einstein equations in vacuum, $E_{\mu\nu}\equiv R_{\mu\nu}- \frac{1}{2} g_{\mu\nu} R=0$. In order to avoid any confusion, we note that the greek indices $\alpha, \beta$ used from \eqref{Lag} onwards, are neither space-time nor the internal indices utilized in \eqref{lineel} and \eqref{sigmarel}. They are just used here to denote the configuration space variables as components of $q=(\Omega,\beta_1,\beta_2)$. We may now proceed to the Hamiltonian description of the model and the derivation of the nonlocal conserved charges.

\section{Phase space formalism and integrability} \label{sec3}

Due to the fact that the Lagrangian \eqref{Lag} is singular i.e. the corresponding Hessian matrix is not invertible, we have to put in use the Dirac-Bergmann algorithm \cite{Dirac,Berg} in order to be led to the Hamiltonian function for the given problem. The corresponding total Hamiltonian is
\begin{equation} \label{HamT}
  H_T = N \mathcal{H} +u_N p_N
\end{equation}
where
\begin{align}
  p_N & \approx 0 \\ \label{qcon}
  \mathcal{H} & = \frac{1}{2} G^{\alpha\beta} p_\alpha p_\beta + V(q) \approx 0
\end{align}
are the two first class constraints of the system, while $u_N$ is an arbitrary function. The ``$\approx$" symbol denotes a weak equality, that is the $p_N$ and $\mathcal{H}$ are zero themselves, but their phase-space gradients are not.

For systems characterized by \eqref{HamT}-\eqref{qcon}, it can be seen that any conformal Killing vector, $\xi=\xi^\alpha \frac{\partial}{\partial q^\alpha}$, of the minisuperspace metric $G_{\alpha\beta}$, i.e. $\mathcal{L}_\xi G_{\alpha\beta} = \omega(q) G_{\alpha\beta}$ generates a (generally) nonlocal conserved charge of the form
\begin{equation} \label{intofmo}
  Q = \xi^\alpha p_\alpha + \int\!\! N(t) \left[\omega(q(t))+F(q(t))\right] V(q(t)) dt,
\end{equation}
where $p_\alpha := \frac{\partial L}{\partial \dot{q}^\alpha}$ are the momenta of the system and $F(q):= \frac{1}{V(q)} \xi^\alpha \partial_\alpha V(q)$. Whenever $\omega(q)=-F(q)$ we obtain a typical linear in the momenta integral of motion $Q=\xi^\alpha p_\alpha$. In any other case there exists a nonlocal part as seen in \eqref{intofmo} owed to the presence of an integral in time over phase space functions. It is easy to check that
\begin{equation}
  \frac{d Q}{d t} = \frac{\partial Q}{\partial t} + \{Q, H_T\} = N (\omega + F) V + N \omega G^{\alpha\beta} p_\alpha p_\beta - N F V = N \omega \mathcal{H} \approx 0 .
\end{equation}
This is a generalization of Kucha\v{r}'s conditional symmetries \cite{Kuchar}. Kucha\v{r} had defined as conditional symmetries, quantities that are linear in the momenta and which have the property of weakly commuting with the Hamiltonian constraint, i.e. $Q=\xi^\alpha p_\alpha$ with $\{Q,\mathcal{H}\} \approx 0 \Rightarrow \{Q,\mathcal{H}\} = s(q) \mathcal{H}$. Here, our $Q$ in \eqref{intofmo}, may additionally have an explicit dependence in time in terms of an integral of phase space functions. The emerging quantities $Q$ are conserved on the constraint surface $\mathcal{H}\approx 0$ and, hence, their existence is tied to the parametrization invariance of the system. In other words, these integrals of motion would not appear in the study of a gauge fixed version of \eqref{Lag}, e.g. if we had considered $N=1$ in the latter (thus missing the quadratic constraint).

For the given minisuperspace metric \eqref{min} it can be seen that there exist ten conformal Killing fields
\begin{equation} \label{CKVs}
  \begin{split}
    & \xi_1 = \partial_\Omega, \quad \xi_2 = \partial_{\beta_1}, \quad \xi_3 = \partial_{\beta_2}, \quad \xi_4 = \Omega \partial_\Omega + \beta_1 \partial_{\beta_1} + \beta_2 \partial_{\beta_2} \\
    & \xi_5 = \beta_1 \partial_\Omega + \Omega \partial_{\beta_1}, \quad \xi_6 = \beta_2 \partial_\Omega + \Omega \partial_{\beta_2}, \quad \xi_7 = \beta_2 \partial_{\beta_1} - \beta_1 \partial_{\beta_2} \\
    &\xi_8 = \frac{1}{2} \left( \beta_1^2 +\beta_2^2 + \Omega^2 \right) \partial_\Omega + \beta_1 \Omega \partial_{\beta_1} + \beta_2 \Omega \partial_{\beta_2}\\
    & \xi_9 = \beta_1 \Omega \partial_{\Omega} + \frac{1}{2} \left( \beta_1^2 - \beta_2^2 + \Omega^2\right) \partial_{\beta_1} + \beta_1 \beta_2 \partial_{\beta_2}, \\
    & \xi_{10} = \beta_2 \Omega \partial_{\Omega} + \beta_1 \beta_2 \partial_{\beta_1} +\frac{1}{2} \left( \Omega^2 - \beta_1^2 +\beta_2^2 \right) \partial_{\beta_2}
  \end{split}
\end{equation}
with the corresponding conformal factors
\begin{equation}
  \begin{split}
    \omega_1 = -3, \quad \omega_2=\omega_3=\omega_7=0, \quad \omega_4 = 2-3\Omega, \quad \omega_5 = -3 \beta_1, \quad \omega_6 = -3\beta_2,\\
     \omega_8 = \frac{1}{2} \left(-3 \beta_1^2-3 \beta_2^2+\Omega  (4-3 \Omega )\right), \quad \omega_9 = \beta_1 (2-3 \Omega ), \quad \omega_{10} = \beta_2 (2-3 \Omega ) .
  \end{split}
\end{equation}
From the latter we see that $\xi_2$, $\xi_3$ and $\xi_7$ are Killing vector fields of $G_{\alpha\beta}$, while $\xi_1$ is a homothetic vector.

The above $\xi_I$s, $I=1,...,10$, can be used to construct ten conserved charges of the form \eqref{intofmo}. The ten functions $F_I$ inside the integral of \eqref{intofmo} can be easily calculated through the relations  $F_I= \frac{1}{V}\xi_I^\alpha \partial_\alpha V$. It is easy to verify that $\frac{dQ_I}{dt}=0$ whenever the Euler-Lagrange equations of \eqref{Lag} are satisfied (of course counting the constraint equation in them). Note that if we had considered as possible candidates for an integral of motion only quantities which are strictly functions of position and momenta, for example $Q= \xi^\alpha p_\alpha$, then this would result in no conserved charge for the given system. This is owed to the fact that the necessary condition $\omega(q)=-F(q)$ for such a conserved charge in \eqref{intofmo} is not satisfied for any of the $\xi_I$ appearing in \eqref{CKVs}. It is due to the nonlocal part that these symmetries can define integrals of motion.

Now that we used the parametrization invariance to derive all linear in the momenta symmetries of the system we can proceed by fixing the gauge. At the same time, and in order to check for the integrability of the system when non-autonomous integrals of motion are present, we promote the time variable to a dynamical degree of freedom; as also happens for regular systems whenever there appears an explicit dependence in time (see \cite{Bouq} and references therein). We first introduce the additional gauge fixing constraint
\begin{equation}
  \chi = N - f(t) \approx 0,
\end{equation}
where $f(t)$ is some appropriate function of the time variable ($f$ can also be a constant). Then, we consider the Hamiltonian
\begin{equation} \label{Hambig}
  H = p_t + H_T + u_\chi \chi,
\end{equation}
with $p_t$ being the canonical conjugate to the new degree of freedom $t$, i.e. $\{t,p_t\}=1$. The constraints $p_N\approx 0$, $\chi \approx 0$ turn into second class since $\{\chi,p_N\}=1$, while the $\mathcal{H}\approx 0$ remains first class, on account of $\{\mathcal{H},p_N\}\approx 0$ and $\{\mathcal{H},\chi\}\approx 0$. The consistency conditions $\dot{p}_N= \{p_N,H\} \approx 0$ and $\dot{\chi}=\{\chi,H\}\approx 0$ lead to the determination of the ``velocities" $u_\chi$ and $u_N$ respectively. We straightforwardly obtain $u_\chi\approx 0$ and $u_N\approx \dot{f}$. As a result, the Hamiltonian \eqref{Hambig} finally reads
 \begin{equation} \label{Hambig2}
  H = p_t + N \mathcal{H} +\dot{f} p_N.
\end{equation}
The Dirac brackets are defined as
\begin{equation}
  \{F,G\}_D = \{F,G\} - \{F,p_N\}\{\chi, G\} + \{F,\chi\}\{p_N, G\} .
\end{equation}
and by considering the ten integrals of motion
\begin{equation}\label{int2}
  Q_I = \xi^\alpha_I p_\alpha + A_I(t), \quad I=1,...,10,
\end{equation}
where
\begin{equation}
  A_I(t) = \int\!\! f(t) \left[\omega_I(q(t))+F_I(q(t))\right] V(q(t)) dt,
\end{equation}
we can see that
\begin{equation}\label{comrel}
  \{Q_I, H\}_D  \approx 0 , \quad \{\mathcal{H} , H\}_D= 0 , \quad \{Q_I, \mathcal{H}\}_D  \neq 0.
\end{equation}
From the last relation we may observe that these ten integrals of motion, even though they commute (weakly) with the Hamiltonian $H$, they do not commute with the quadratic constraint $\mathcal{H}\approx 0$. The reduced (from the second class constraints) phase space is eight dimensional and spanned by $t$, $\Omega$, $\beta_1$, $\beta_2$ and their conjugate momenta. The corresponding reduced Hamiltonian is
\begin{equation}
  H_{\text{red}} = p_t + f(t) \mathcal{H} .
\end{equation}
In order to talk about Liouville integrability we need four independent phase space functions that are in involution. From the ten conformal Killing vectors in \eqref{CKVs} we may notice that there exist two three dimensional Abelian subalgebras: the first involving $\xi_1$, $\xi_2$ and $\xi_3$ and the second consisting of $\xi_8$, $\xi_9$ and $\xi_{10}$. This results in two Abelian three dimensional Poisson algebras in the corresponding $Q_I$'s. If we also consider, according to the first of \eqref{comrel}, that all the $Q_I$'s have the property of weakly commuting with $H$, then we have two choices for a set of four independent integrals of motion that are in involution
\begin{equation}
  \begin{split}
   \{Q_I,H\}_D = \{Q_I, H_{\text{red}}\}\approx 0, \quad & \{Q_I,Q_J\}_D=\{Q_I,Q_J\} =0 , \\
   & I,J=1,2,3, \; \text{or} \; I,J=8,9,10. \\
  \end{split}
\end{equation}

We have to make the following observations:
\begin{itemize}
  \item The set of four mutually commuting phase space functions exists only on the constraint surface $\mathcal{H}\approx 0$. We see that $\{Q_I,H_{\text{red}}\}\approx 0$, i.e. it is a weak equality, which practically means that the system may be characterized as integrable only because of the zero value of the Hamiltonian constraint.

  \item We may be aware of the existence of the ten integrals $Q_I$, but their explicit dependence on $t$ cannot be known; not unless we have the solution in terms of the three $q(t)$. However, the gauge fixing condition may be applied in such a manner that, at least for one of the $Q_I$'s, the corresponding function $A_I(t)$ becomes apparent. For example, if we choose in, say $Q_1$, the lapse function to be $N(t)=f(t)= \left[ \left(\omega_1(q)+F_1(q)\right)V(q) \right]^{-1}$, then in this gauge, the conserved charge reads $Q_1=\xi_1^\alpha p_\alpha + t=$const. The rest of the $A_I(t)$ functions however remain unknown as long as we do not have the explicit expressions for $\Omega$, $\beta_1$ and $\beta_2$ as functions of $t$ that solve the equations of motion. This in itself is an important result since we have at our disposal some specific gauge choices under which a second integral of motion independent of the Hamiltonian can have an analytic form in phase space variables and $t$. Previous theorems excluded the existence of such a quantity \cite{non1,non2}; however, they were restricted to consider functions only in the positions and the momenta, but not in time. We see that by allowing explicit time dependence in the conserved charges we are able to reveal such quantities.

  \item A natural question to ask at this point would be the possible physical significance of the constants of integration corresponding to the $Q_I$'s as they have emerged from our treatment. This is a highly non trivial inquiry, even in the case where conserved charges of a local form are involved in a mini-superspace analysis. This is owed to the fact that the mini-superspace system is not sensitive to three-dimensional spatial diffeomorphisms of the base manifold metric \eqref{lineel}. In order to be able to distinguish which combination of constants of integration is of physical importance we would need to have at our disposal the analytic solution, insert it into line element \eqref{lineel} and then check if there exist any spatial diffeomorphisms that further reduce the number of these constants. The physically relevant constants (or the appropriate combinations of them) would be those that are essential for the space-time geometry and which cannot be absorbed by any diffeomorphisms. A particular example where four constants of integration emerging from conserved mini-superspace charges are reduced to just two that are physically relevant can be seen in the classical part of the analysis performed in \cite{RNpaper}.

  \item Of course, one may wonder whether a solution $q(t)$ exists in the first place, so that there is a sense in considering the functions $A_I(t)$ in the integrals of motion \eqref{int2}. This is something that we immediately check in the paragraph that follows.
\end{itemize}

It can be easily seen that in the gauge $N= 12e^{-3\Omega}$, the three integrals of motion $Q_1$, $Q_2$ and $Q_3$ are respectively (when the momenta are substituted with respect to velocities):
\begin{equation} \label{threeint}
  \dot{\Omega} = \int \!\! B_1(t) dt, \quad  \dot{\beta}_1 = \int \!\! B_2(t) dt, \quad \dot{\beta}_2 = \int \!\! B_3(t) dt ,
\end{equation}
where
\begin{align}
 B_1=& 24\Big(2 e^{6 \left(\beta_1+\sqrt{3} \beta_2\right)} + 2 e^{6 \beta_1+2 \sqrt{3} \beta_2} +2 e^{12 \beta_1+4 \sqrt{3} \beta_2} - e^{12 \beta_1+8 \sqrt{3} \beta_2} \nonumber \\
 & -e^{12 \beta_1} - e^{4 \sqrt{3} \beta_2}\Big) e^{-4 \left(2 \beta_1+\sqrt{3} \beta_2+\Omega \right)} \\
 B_2 =& 24\Big(2 e^{12 \beta_1+4 \sqrt{3} \beta_2} - e^{6 \left(\beta_1+\sqrt{3} \beta_2\right)} - e^{6 \beta_1+2 \sqrt{3} \beta_2} - e^{12 \beta_1+8 \sqrt{3} \beta_2} \nonumber\\
 & -e^{12 \beta_1} + 2 e^{4 \sqrt{3} \beta_2}\Big) e^{-4 \left(2 \beta_1+\sqrt{3} \beta_2+\Omega \right)}\\
 B_3 =& 24\sqrt{3} \left(1-e^{4 \sqrt{3} \beta_2}\right) \left(e^{6 \beta_1+4 \sqrt{3} \beta_2}+e^{6 \beta_1}-e^{2 \sqrt{3} \beta_2}\right) e^{-2 \left(\beta_1+2 \sqrt{3} \beta_2+ 2\Omega \right)}.
\end{align}
The total derivatives of \eqref{threeint} with respect to the time $t$ lead to the spatial equations of the system
\begin{equation} \label{spaeq}
  \ddot{\Omega} = B_1(\Omega,\beta_1\beta_2), \quad \ddot{\beta}_1 = B_2 (\Omega,\beta_1\beta_2), \quad \ddot{\beta}_2 = B_3 (\Omega,\beta_1\beta_2).
\end{equation}
Substitution of the latter in the Euler-Lagrange equations of \eqref{Lag} for $\Omega$, $\beta_1$ and $\beta_2$ results in the satisfaction of the latter two, while the first becomes proportional to the constraint equation $\frac{\partial L}{\partial N}=0$. We may observe that equations \eqref{spaeq} do not include first order derivatives of the configuration space variables $q=(\Omega,\beta_1,\beta_2)$. It is known that the Bianchi type IX equations can be brought to this form. Here, we see that this property is a consequence of the existing nonlocal conserved charges. By writing the first order equivalent system of \eqref{spaeq} it is obvious that it satisfies Peano's theorem, thus at least one solution always exists inside a given domain of definition and given initial conditions. So we can say that there is a sense in considering the $A_I(t)$ in \eqref{int2} even though we are not aware of their explicit dependence in $t$.

As we demonstrated, the sufficient number of mutually commuting integrals of motion exists and thus the system can be characterized - at least locally - as Liouville integrable.

\section{A few comments from the theory of constrained systems and the importance of $\mathcal{H}\approx 0$} \label{sec4}

It may be stated that the result obtained through our proof in the previous section is in contradiction with the implied chaotic behaviour of the model from various previous works. Although we are not experts in chaos theory, we believe that such a comparison is neither trivial nor straightforward. A first issue is that - unlike Liouville integrability - chaos as a notion is not based on a universally used, exact mathematical definition; it is rather a qualitative behaviour of a system. What is more, it is not obvious that the methods of examining a chaotic behaviour in regular mechanical systems can be directly used in the singular case without appropriate modifications. To this end let us just review some basic facts of the mixmaster model that - not always, but many times - are overlooked in the literature as trivial and which we consider that here play an important role.

The point of view is commonly encountered, according to which the Hamiltonian constraint $\mathcal{H}\approx 0$ can be thought of as an integral of motion which for some reason is restricted to the value zero. In other words that $\mathcal{H}$ is the Hamiltonian of a system of three degrees of freedom, where the \emph{ad hoc} condition $\mathcal{H}=0$ has been set. This often creates the misconception that one has to deal with three independent degrees of freedom where $\mathcal{H}=0$ sets a restriction upon the six expected constants of integration emerging from the general solution of the spatial equations. However, the effect of $\mathcal{H}\approx 0$ has much more intriguing consequences than that. Technically, $\mathcal{H}$ is not an integral of motion for the system (an integral of motion can in general assume arbitrary constant values on mass shell). The constraint equation $\mathcal{H}\approx 0$ is a self consistency condition so that the Hamiltonian formalism, with the Hamiltonian function $H_T$ in \eqref{HamT}, corresponds to the parametrization invariant Lagrangian \eqref{Lag} that describes the original system. If we translate $\mathcal{H}\approx 0$ in velocity phase space coordinates, we see that it is the Euler-Lagrange equation for the degree of freedom $N$, i.e. $\frac{\partial L}{\partial N}=0$, to be considered on an equal footing with the spatial Euler-Lagrange equations for $\Omega,\beta_1$ and $\beta_2$. As a result, the existence of the constraint does not imply the removal of a single combination of the integration constants appearing in the general solution, but rather the removal of a full degree of freedom among the three $\Omega,\beta_1,\beta_2$. This means that the mixmaster model exhibits only two independent (or physical) degrees of freedom with the third expressing just a gauge choice. For the mathematical formula that calculates the number of physical degrees of freedom in constrained systems we can refer the interested reader to the textbook \cite{Teit} and for an equivalent counting in the Lagrangian formalism in \cite{DHM}.

To demonstrate the above mentioned fact, consider that you solve the constraint equation $\frac{\partial L}{\partial N}=0$ algebraically with respect to the lapse $N$. Substitution of this result into the three spacial equations results into having only two of the latter being independent. Thus, the system obtained consists of two different, second order, coupled ordinary differential equations for three functions the $\Omega,\beta_1,\beta_2$. One of them can be considered as a time variable  since, up to now, the gauge is not fixed; thus, the ensuing general solution is expected to  contain only four constants of integration. What is more, which  two of the three degree of freedom we may consider as the physical ones and which as the gauge, is completely at our disposal. In principle, it makes no difference for the system. What we want to demonstrate by this is that a method of analysis respecting parametrization invariance must be insensitive to these choices. The physically important results cannot depend neither on which two functions of the triplet $(\Omega, \beta_1,  \beta_2)$ one may consider as physical, nor on the explicit dependence in $t$ generated by the third function. In our analysis, we have respected this principle by treating the system in the complete phase space on which $H_T$ is defined and proceeded following exactly the theory of constrained systems. In view of this, let us  note  that conclusions about ergodicity, based on plots of $\beta_1/\Omega$ vs $\beta_2/\Omega$ etc., are not straightforward; before they can be reached, further and rather careful investigations are needed.

Another point we want to make is the importance of $\mathcal{H}\approx 0$ for the existence of the conserved charges that we defined. The aforementioned quantities are conserved only on the constraint surface. This - unlike what happens in the case of regular systems where the Hamiltonian may assume arbitrary constant values - probably makes the integrability of the system extremely sensitive to the satisfaction of the $\mathcal{H}\approx 0$ condition. It has been noticed in the literature that this is always an issue when numerical methods are being employed  \cite{Hobill,Uggla} or other types of approximations like involving the Kasner map conjecture \cite{Uggla}. In our work we follow an exact treatment in phase space (having emerged from the treatment of the original parametrization invariant Lagrangian) without making a single approximation throughout our analysis. In our view, this is what allowed the revealing of the conserved quantities which make the difference towards the (local) integrability of the model.

\section{Conclusions}

We have investigated the Liouville integrability of the anisotropic Bianchi type IX model in vacuum. By allowing functions that have an explicit time dependence in terms of an integral of phase space functions, we obtained a generalized class of conserved charges. The latter are constants of motion strictly on the constrained surface $\mathcal{H}\approx 0$ since they weakly commute with the Hamiltonian $H$. From the ten derived quantities whose dependence in the momenta is linear, there exist two three dimensional Abelian Poisson algebras that together with the extended in $t$ Hamiltonian in \eqref{Hambig2} form two sets of four first integrals in involution for the system. Even though the existence of these sets is guaranteed by the fact that locally a solution $\Omega(t)$, $\beta_1(t)$, $\beta_2(t)$ always exists, the explicit dependence of each integral in $t$ cannot be known - at least not for all the nonlocal charges. An appropriate gauge choice can be made so that one of the nonlocal integrals of motion assumes a local form, but for the rest their explicit dependence in $t$ remains unknown. For the physical interpretation of these constants of motion one has to take into account the limitations of the mini-superspace analysis. The fact that spatial diffeomorphisms - which are hindered at the minisuperspace level - can be used in the base manifold metric to absorbe some of these constants. Thus, without the explicit solution, it is not clear which combination of them may be essential for the space-time geometry.

Lastly, we would like to notice that the same conclusion about Liouville integrability can be reached if we include a cosmological constant term or any other matter content that does not affect the minisuperspace metric \eqref{min}. Since the existence of the two sets of commuting conformal Killing fields is not affected. In this situation only the functions $F_I$, which are associated to the Lie derivative of the $\xi_I$ on the minisuperspace potential change. Of course, the potential needs to be smooth enough for solutions to exist in some domain of definition for the $\Omega$, $\beta_1$ and $\beta_2$.


\begin{thebibliography}{99}

\bibitem{Cushman} R. Cushman and J. \'{S}niatycki, \textit{Reports in Math. Phys.} \textbf{36}, (1995) 75

\bibitem{Intp1} G. Contopoulos, B. Grammaticos and A. Ramani, \textit{J. Phys. A: Math. Gen.} \textbf{26}, (1993) 5795

\bibitem{Cotsakis} S. Cotsakis, P. G. L. Leach, \textit{J. Phys. A: Math. Gen.} \textbf{27}, (1994) 1625

\bibitem{Intp2} F. Christiansen, H. H. Rugh and S. E. Rugh, \textit{J. Phys. A: Math. Gen.} \textbf{28}, (1995) 657

\bibitem{Intp3} G. Contopoulos, B. Grammaticos and A. Ramani, \textit{J. Phys. A: Math. Gen.} \textbf{27}, (1994) 5357

\bibitem{Intp4} A. Latifi, M. Musette and R. Conte, \textit{Phys. Lett. A} \textbf{194}, (1994) 83

\bibitem{Intp5} K. Andriopoulos and P. G. L. Leach, \textit{J. Phys. A: Math. Theor.} \textbf{41}, (2008) 155201

\bibitem{Barrow} J. D. Barrow, \textit{Phys. Rep.} \textbf{85}, (1982) 1

\bibitem{chaos1} D. F. Chernoff and J. D. Barrow, \textit{Phys. Rev. Lett.} \textbf{50}, (1983) 134


\bibitem{chaos3} I. M. Khalatnikov, E. M. Lifshitz, K. M. Khanin, L. N. Shehur and Ya. G. Sinai, \textit{J. Stat. Phys.} \textbf{38}, (1985) 97

\bibitem{chaos2} M. H. Bugalho, A. Rica da Silva and J. S. Ramos, \textit{Gen. Rel. Grav.} \textbf{18}, (1986) 1263

\bibitem{Matsas2} G. Francisco and G. E. A. Matsas, \textit{Gen. Relativ. Gravit.} \textbf{20}, (1988) 1047

\bibitem{Ellis} A. B. Burd, N. Buric and G. F. R. Ellis, \textit{Gen. Relativ. Gravit.} \textbf{22}, (1990) 349

\bibitem{Rugh}  S. E. Rugh and  B. J. T Jones, \textit{Phys. Lett. A} \textbf{147}, (1990) 353

\bibitem{Hobill} D. Hobill, D. Bernstein, M. Welge and D. Simkins, \textit{Clas. Quantum Grav.} \textbf{8}, (1991) 1155

\bibitem{Berger1} B. K. Berger, \textit{Gen. Rel. Grav.} \textbf{23}, (1991) 1385

\bibitem{Matsas1} K. Ferraz, G. Francisco and G. E. A. Matsas, \textit{Phys. Lett. A} \textbf{156}, (1991) 407

\bibitem{Pullin} J. Pullin, \textit{Time and Chaos in General Relativity}, in \textit{SILARG VII Relativity and Gravitation: Classical and Quantum}, World Scientific, Singapore, (1991)

\bibitem{chaos4.0} P. K.-H. Ma and J. Wainwright \textit{A Dynamical Systems Approach to the Oscillatory Singularity in Bianchi Cosmologies}, in  \textit{Deterministic Chaos in General Relativity}, eds. D. W. Hobill, A. Burd, and A. Coley, Plenum, New York, (1994)

\bibitem{chaos4} N. J. Cornish and J. J. Levin, \textit{Phys. Rev. Lett.} \textbf{78}, (1997) 998

\bibitem{chaos5} N. J. Cornish and J. J. Levin, \textit{Phys. Rev. D} \textbf{55}, (1997) 7489

\bibitem{chaos6} S. E. Rugh, \textit{Chaos in the Einstein Equations-Characterization and Importance}, in \textit{Deterministic Chaos in General Relativity}, eds. D. W. Hobill, A. Burd, and A. Coley, Plenum, New York, (1994)

\bibitem{chaos7} G. Contopoulos, N. Voglis and C. Efthymiopoulos, \textit{Chaos in Relativity and Cosmology}, in \textit{Impact of Modern Dynamics in Astronomy}, eds. J. Henrard and S. Ferraz-Mello, Springer Science+Business Media,  Dordrecht (1999)

\bibitem{Motter} A. E. Motter and P. S. Letelier, \textit{Phys. Lett. A} \textbf{285}, (2001) 127

\bibitem{Arnold} V. I. Arnold, \textit{Mathematical Methods of Classical Mechanics}, 2nd Edition, Springer-Verlag, New York, Berlin, Heidelberg (1989)

\bibitem{Painleve} R. Conte and M. Musette, \textit{The Painlev\'e Handbook}, Springer Netherlands (2008)

\bibitem{Ryan} M. P. Ryan Jr. and L. C. Shepley, \textit{Homogeneous Relativistic Cosmologies}, Princeton Series in Physics, Princeton, New Jersey 1975

\bibitem{Jan} R. T. Jantzen, \textit{Spatially Homogeneous Dynamics: A Unified Picture}, in \textit{Gamov Cosmology}, eds. R. Rufini, and F. Melchiorri, North Holland, Amsterdam (1987)

\bibitem{Misner} C. W. Misner, \textit{Phys. Rev. Lett.} \textbf{22}, (1969) 1071

\bibitem{BKL1} V. A. Belinskii, I. M. Khalatnikov and E. M. Lifshitz, \textit{Advances in Physics} \textbf{19}, (1970) 525

\bibitem{BKL1b} V. A. Belinskii,  E. M. Lifshitz and I. M. Khalatnikov, \textit{Sov. Phys. JETP} \textbf{33}, (1971) 1061

\bibitem{BKL2} V. A. Belinskii, I. M. Khalatnikov and E. M. Lifshitz, \textit{Advances in Physics} \textbf{31}, (1982) 639

\bibitem{Ryan0} A. R. Moser, R. A. Matzner and M. P. Ryan Jr., \textit{Ann. Phys.} \textbf{79}, (1973) 558

\bibitem{Berger2} B. K. Berger, \textit{Phys. Rev. D} \textbf{49}, (1994) 1120

\bibitem{Berger3} B. K. Berger, D. Garfinkle and E. Strasser, \textit{Class. Quantum Grav.} \textbf{14}, (1997) L29-L36

\bibitem{Ringstrom} H. Ringstr\"om, \textit{Ann. Henri Poincar\'e} \textbf{2}, (2001) 405

\bibitem{Uggla} J. M. Heinzle and C. Uggla, \textit{Clas. Quantum Grav.} \textbf{26} (2009) 075016

\bibitem{WHsu} J. Wainwright and L. Hsu, \textit{Clas. Quantum Grav.} \textbf{6} (1989) 1409

\bibitem{LeBlanc1} V. G. LeBlanc, D. Kerr, J. Wainwright, \textit{Clas. Quantum Grav.} \textbf{12} (1995) 513

\bibitem{Bergerf} B. K. Berger, \textit{Clas. Quantum Grav.} \textbf{13} (1996) 1273

\bibitem{LeBlanc2} V. G. LeBlanc, \textit{Clas. Quantum Grav.} \textbf{14} (1997) 2281

\bibitem{Burdrev} A. Burd, \textit{How Can You Tell If the Bianchi IX Models Are Chaotic?}, in  \textit{Deterministic Chaos in General Relativity}, eds. D. W. Hobill, A. Burd, and A. Coley, Plenum, New York, (1994)

\bibitem{non1} A. J. Maciejewski and Marek Szyd{\l}owski, \textit{J. Phys. A: Math. Gen.} \textbf{31}, (1998) 2031

\bibitem{non2} J. Llibre and C. Valls, \textit{J. Math. Phys.} \textbf{47}, (2006) 022704

\bibitem{tchris} T. Christodoulakis, N. Dimakis and Petros A. Terzis, \textit{J. Phys. A} \textbf{47}, (2014) 095202

\bibitem{Kuchar} K. V. Kucha\v{r}, \textit{J. Math. Phys.} \textbf{23}, (1982) 1647

\bibitem{Dirac} P. A. M. Dirac, \textit{Lectures on Quantum Mechanics}, Yeshiva University, Academic Press, New York, 1964

\bibitem{Berg} J. L. Anderson and P. G. Bergmann, \textit{Phys. Rev.} \textbf{83}, (1951) 1018

\bibitem{Bouq} S. Bouquet and A. Bourdier, \textit{Phys. Rev. E} \textbf{57}, (1998) 1273

\bibitem{RNpaper} N. Dimakis, A. Karagiorgos, T. Pailas, Petros A. Terzis and T. Christodoulakis, \textit{Phys. Rev. D} \textbf{95}, (2017) 086016

\bibitem{Teit} M. Henneaux and C. Teitelboim, \textit{Quantization of Gauge Systems}, Princeton University Press, Princeton, New Jersey (1991)

\bibitem{DHM} B. D\'iaz, D. Higuita and M. Montesinos, \textit{J. Math. Phys.} \textbf{55}, (2014) 122901

\end{thebibliography}
\end{document}